# Characterization of physicochemical and colloidal properties of hydrogel chitosan-coated iron-oxide nanoparticles for cancer therapy

E Catalano[1], A Di Benedetto[2]

[1] Department of Clinical Molecular Biology (EpiGen), Akershus University Hospital, University of Oslo (UiO), Oslo, Norway. [2] Dipartimento di Fisica G. Occhialini, Università Milano-Bicocca, Piazza della Scienza, 3, 20126 Milano, Italia

E-mail: enrico.catalano@medisin.uio.no

**Abstract**. Superparamagnetic iron oxide nanoparticles have recently been investigated for their potential to kill cancer cells with promising results, owing to their ability to be targeted and heated by magnetic fields. In this study, novel hydrogel, chitosan $Fe_3O_4$ magnetic nanoparticles were synthesized to induce magnetic hyperthermia, and targeted delivering of chemotherapeutics in the cancer microenvironment. The characteristic properties of synthesized bare and CS-MNPs were analyzed by various analytical methods: X-ray diffraction, Fourier transformed infrared spectroscopy, Scanning electron microscopy and Thermo-gravimetric analysis/differential thermal analysis. Magnetic nanoparticles were successfully synthesized using the co-precipitation method. This synthesis technique resulted in nanoparticles with an average particle size of 16 nm. The pure obtained nanoparticles were then successfully encapsulated with 4-nm-thick chitosan coating. The formation of chitosan on the surface of nanoparticles was confirmed by physicochemical analyses. Heating experiments at safe magnetic field (f = 100 kHz, H =10-20 kA $m^{-1}$) revealed that the maximum achieved temperature of water stable chitosan-coated nanoparticles (50 mg $ml^{-1}$) is fully in agreement with cancer therapy and biomedical applications.

## 1. Introduction

Magnetic nanomaterial-based systems are exploited to develop several novel therapeutic opportunities for cancer therapy based merely upon their unique physicochemical and magnetic properties [1]. Nowadays, magnetic nanoparticles (MNPs), due to their particular physicochemical properties, have attracted growing interest in biomedicine and bioengineering such as magnetically triggered drug delivery [2], bio-separation [1], magnetic resonance imaging [1], bio-macromolecule purification, and magnetic hyperthermia [3, 4]. Magnetic hyperthermia has long been used as a treatment option for cancer [1]. In this method MNPs are introduced to the tumor tissue and then exposed to an alternating magnetic field (AMF). This method leads to heat generation by the MNPs, that can kill the cancerous cells [3, 4]. In magnetically triggered drug delivery systems, MNPs are normally coated with a thermosensitive polymer as drug carrier. When these MNPs are exposed to an AMF, due to the heat generated by magnetic core, phase transition occurs in the thermosensitive polymer and drug is released [1]. The most common MNPs used for magnetic hyperthermia or drug delivery are $Fe_3O_4$ or

γ-$Fe_2O_3$ [5]. Chitosan, as a multifunctional natural polymer containing large numbers of active amino groups together with hydroxyl groups in the backbone, can interact with mucins and open the tight junctions between epithelial cells, which are in favor of the drug transport. The chitosan-based delivery systems have shown great potential to deliver nucleic acids, therapeutic proteins, vaccinations and other bio-molecules [6].

## 2. Materials and methods
*2.1 Synthesis of Iron Oxide Nanoparticles*
Magnetic iron oxide nanoparticles (MNPs) were prepared by alkaline co-precipitation of ferrous chloride tetrahydrate, $FeCl_2 \cdot 4H_2O$ (1.34 g) and ferric chloride hexahydrate $FeCl_3 \cdot 6H_2O$ (3.40 g) at 1:2 ratio. The salts were dissolved in 150 mL deionized water within a three necked glass balloon. The glass balloon was placed in a heating mantle and stirred with a magnetic stirrer. It was vigorously stirred at 90ºC in the presence of $N_2$ gas. Ammonium hydroxide ($NH_4OH$) was added to the system drop wise. The process was ended by washing with deionized water until the solution pH was 9.0. The solution was then centrifuged at 5000 rpm for 30 minutes. The precipitates were collected and dried in the incubator at 55ºC. The black precipitates were then turned into brown.

*2.2 Synthesis of Chitosan-coated Magnetic Iron Oxide Nanoparticles*
Chitosan-coated magnetic iron oxide nanoparticles (CS-MNPs) were in situ synthesized by the co-precipitation of Fe (II) and Fe (III) salts in the presence of chitosan and trisodium phosphate molecules. Chitosan was previously prepared with degree of deacetylation 75% by titrimetric method [7]. Trisodium phosphate was used for the crosslinking of low molecular weight chitosan polymers. Chitosan (0.15 g) was dissolved in 30 ml of 1% acetic acid and the pH was adjusted to 4.8 by 10M NaOH. Iron salts (1.34 g $FeCl_2 \cdot 4H_2O$ and 3.40 g $FeCl_3 \cdot 6H_2O$) were dissolved in 30 ml of 0.5% chitosan solution. The solution was then vigorously stirred at 2000 rpm. 10 ml of 22.5% trisodium phosphate and different amounts of 32% $NH_4OH$ (18, 20, 22, 25 mL) were added to the solution to obtain the final $NH_4OH$ concentration of 31%, at room temperature. The ammonia solution was added very slowly to produce smaller sized nanoparticles. The resulting solution was stirred for an additional 1 hour. The colloidal chitosan coated magnetic $Fe_3O_4$ nanoparticles were extensively washed (3 times) with deionized water and separated by centrifugation and drying.

*2.3 X-ray diffraction (XRD)*
X-ray diffraction (XRD) was used to investigate the mineral phase of obtained materials. XRD measurements were performed using a D8 ADVANC diffractometer (BRUKER AXS GmbH, Germany) with Cu Kα radiation (λ = 1.54178 Å), with the operation voltage and current at 40 kV and 40 mA, respectively. The 2θ range was from 5°-90° in steps of 0.02°.

*2.4 Fourier transformed infrared spectroscopy (FTIR)*
The FTIR spectroscopy of the sample was taken in the region between 500-4000 $cm^{-1}$ (with Perkin Elmer 1650, USA) on a Thermo-Nicolet Avatar 370 model FTIR in order to understand the chemical and structural nature of the chitosan coated iron-oxide nanoparticles.

*2.5 Thermo-gravimetric analysis/differential thermal analysis/derivative thermo gravimetric analysis (TG/DTA/DTG)*
TGA shows the change in mass with increase of temperature. This studies were carried out by TG/DTA 6300, SII Nano Technology, Japan by heating the sample at 20ºC /min in the temperature range 100 - 1000ºC in nitrogen atmosphere.

*2.6 Morphological analysis of the CS-MNPs*
The size of the prepared sample (CS-MNPs) was related to properties such as bioavailability and surface modification for in vivo application. The particles were inspected on an FEI Quanta 200 scanning electron microscope (SEM) at an accelerating voltage of 20 kV to identify average diameter, interconnectivity, and any agglomeration.

*2.7 Zeta potential measurements*

The evolution of ζ potential towards pH of the nanoparticles solution was evaluated using a ζ potentiometer (Zetasizer Nano ZS90; Malvern Instruments, Malvern, UK) in order to monitor the stability of nanoparticles in aqueous media. ζ potential measurements were performed both for bare and chitosan coated $Fe_3O_4$ and a comparison was performed.

The nanoparticles-containing solution were diluted 100-fold with deionized water before testing. The analysis was carried out at room temperature, three measurements per sample were performed and the values reported as the average ± SD.

*2.8 Calorimetric measurements*

The heat generation ability of the iron-oxide nanoparticles for biomedical application was investigated using a homemade induction heating instrument that induced an alternating magnetic field in five turns (5 cm diameter) induction coil. A close loop water temperature control was made to keep the temperature of the coil at ambient temperature. The stable suspension of nanoparticles (50 mg ml$^{-1}$) in 1.5 ml plastic microtube, thermally insulated using styrofoam, was placed in the centre of coil and then irradiated by magnetic field (f = 100 kHz, H = 10-20 kA m$^{-1}$). The temperature of the sample was monitored by an alcohol thermometer. The value of the specific absorption rate (SAR) of the sample was calculated according to [Eq. (1)]

$$\text{SAR} = \left(\frac{C_{\text{suspension}}}{X_{\text{NP}}}\right)\left(\frac{dT}{dt}\right) \quad (1)$$

where $C_{\text{suspension}}$ is the specific heat of the suspension [8], C indicates precisely the heat capacity of the fluid per unit mass of nanoparticles and their suspension in solution, $X_{\text{NP}}$ is the weight fraction of nanoparticles ($X_{\text{NP}} = 0.05$) and dT/dt is the initial slope of the temperature curve vs. time.

## 3. Results and discussion

The characteristic properties of synthesized bare and CS-MNPs were analyzed by various analytical methods. Crystal structures of synthesized MNPs were analyzed by XRD. The chemical groups and chemical interactions involved in synthesized MNPs were identified using the FTIR methods. The sizes of magnetic core and morphological properties were observed through SEM. The qualitative and quantitative information about the volatile compounds of CS-MNPs were performed using TG/DTA.

*3.1. Structural and component analysis of CS-MNP nanocarriers*

The crystal structures of $Fe_3O_4$ MNPs and CS-$Fe_3O_4$ MNPs were characterized by X-ray diffraction (XRD), see in Fig. 1(a). No obvious diffraction peak is detected, but only a broad hump at around 2θ = 20.5° in the XRD pattern of chitosan, indicating that raw chitosan has an amorphous structure. Fig. 1a shows the XRD pattern of synthesized crystalline iron oxide ($Fe_3O_4$) nanoparticle. $Fe_3O_4$ exhibits a strong and sharp peak in the diffractogram. The diffraction peaks are obtained at (220), (311), (400), (422), (511), and (440) reflections, which are the characteristic peaks of magnetite ($Fe_3O_4$). The peaks in the XRD patterns of iron oxide are compared with the standards, (JCPDS card file no 77-1545), appeared in both bare and chitosan coated iron oxide nanoparticle. The diffractograms of chitosan are showed in Fig. 1(a)-(c), which revealed the characteristic peak between 9º-20º with comparable degree of crystallinity (Fig. 1a) [9]. The XRD spectrum of chitosan is reported in various works [10, 11] which show peaks of low intensity values around 2θ=10º, and a broad peak at around 2. XRD result exposed a broader peak in chitosan coated iron-oxide nanoparticle (CS-MNPs), which indicates the smaller crystallite size of the produced CS-MNPs than the one of the starting iron oxide nanoparticles. This may be due to the method of preparation of composite that causes the coating of chitosan on iron oxide nanoparticle. The crystallite size of both iron oxide and produced composite (CS-MNPs) were calculated from peaks at 2θ = 35.55º corresponding to iron oxide phase using Scherrer [Eq. (2)]

$$D_p = \frac{0.94\lambda}{\beta_{1/2} cos\theta}$$

(2)

where β is the FWHM of diffraction peak, λ is the wave length of X-ray (0.154 nm), L is the crystallite size, and θ is the Bragg peak position. The crystallite size was found to be 13.4 nm and 11.6 nm for MNPs and CS-MNPs respectively.

*3.2 FT-IR analysis of CS-MNPs*
FT-IR spectral analysis was employed to confirm the introduction of chitosan in $Fe_3O_4$ MNPs and the results are shown in Fig. 1(d). From the spectrum of chitosan, distinctive adsorption peaks appear at 1080 $cm^{-1}$, 1653 $cm^{-1}$ (skeletal vibration involving the C–O stretching) and 2848 $cm^{-1}$ (asymmetric stretching of the C–H bridge) [12]. The absorption peak appears at 3432 $cm^{-1}$ assigned to the axial stretching vibration of O–H, superimposed to the N–H stretching band and inner hydrogen bonds of the polysaccharide, indicating the remaining amino and hydroxyl groups providing the coordination ability with metal [13]. The characteristic absorption peak appears at 580 $cm^{-1}$ attributed to the Fe–O bond, confirming the existence of $Fe_3O_4$ [14]. In comparison with the chitosan spectrum, all the significant peaks of chitosan in the range of 900-1200 $cm^{-1}$ are present in the CS-MNPs with a small shift. Moreover, one main characteristic peak appears at 1635 $cm^{-1}$ due to the bands of COOM (M represents metal ions) groups, indicating the COOH groups of reaction of chitosan with the surface OH groups of $Fe_3O_4$ particles and resulting in the formation of the iron carboxylate [15]. In terms of CS-MNPs, the stretching vibrations of C=N bond at 1640 $cm^{-1}$ could be observed. This peak clearly indicates the formation of the Schiff's base as a result of the reaction between the carbonyl group of glutaraldehyde and the amine group of chitosan chains. Additionally, the C–O absorption peak of secondary hydroxyl group becomes stronger and moves from 1080 $cm^{-1}$ to 1060 $cm^{-1}$. Finally, the absorption of Fe–O bond of $Fe_3O_4$ still exists in the CS-MNPs, which further demonstrates that $Fe_3O_4$ nanoparticles have been wrapped by crosslinked chitosan, and the results are in good agreement with the conclusions of XRD analyses.

*3.3 Scanning Electron Microscopy (SEM) of CS-MNPs*
The typical SEM micrograph for the CS-MNP nanoparticles was shown in Fig. 2(a). From Fig. 2(a), it is clear that $Fe_3O_4$ nanoparticles bonded with chitosan have well-shaped spherical form with smooth surface, and its average particle size is approximately 16±3 nm with a narrow size distribution. The diameters of CS-MNPs are slightly bigger than the sizes of bare MNPs, due to the chitosan polymer coating on the surface of $Fe_3O_4$ particles (thickness of chitosan layer ~ 4 nm). The SEM image can be considered as the persuasive evidence that CS-functionalized MNPs are fabricated successfully. Meanwhile, it displays a slight reunion due to its smaller particle size, high surface energy and surface activity, thus, composite nanoparticles are prone to attract each other to formulate a stable state.

*3.4 Thermal Gravimetric Analysis*
The TGA-FTIR analysis of bare and chitosan coated $Fe_3O_4$ nanoparticles provides qualitative and quantitative information about the volatile components of the nanoparticles. The TGA curve shows that the weight loss of bare MNPs over the temperature range from 50ºC to 800ºC is about 7%. This might be because of the loss of residual water in the sample. The CS-MNPs gave their distinctive TGA curves, which can provide indications of the content of chitosan polymers [16]. The chains of chitosan degrade at about 210ºC and the temperature of decomposition was around 600ºC (Fig. 3).

The results of TGA-DTGA demonstrated that most of the organic layer, chitosan, was removed as $CO_2$ at high temperatures. TGA result indicated that the chitosan content increased at higher temperature and maximum amount is found about 29% by weight.

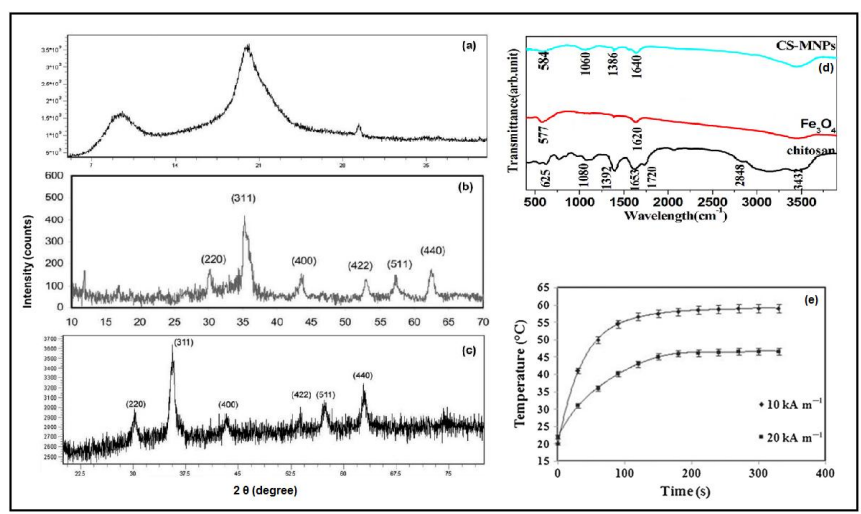

**Figure 1.** XRD pattern of synthesized crystalline iron oxide ($Fe_3O_4$) nanoparticle (a), chitosan (b), Chitosan-coated MNPs (c) and FT-IR spectra of CS-MNP nanoparticles (d), temperature vs. time for aqueous suspension of iron-oxide nanoparticles at different magnetic fields (e).

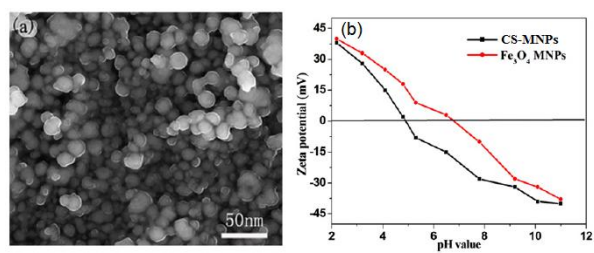

**Figure 2.** SEM images of CS-MNPs (a) and (b) Zeta potentials of bare $Fe_3O_4$ nanoparticles and CS-MNPs at different pH values.

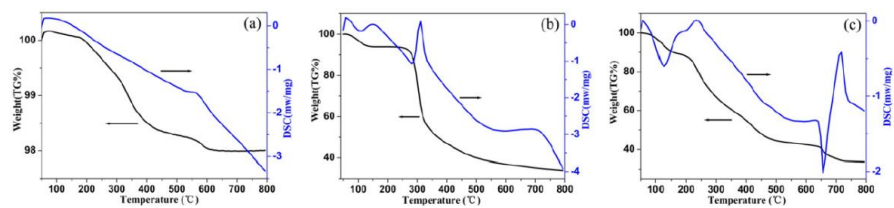

**Figure 3.** Thermogravimetric analysis curves of (a) $Fe_3O_4$ magnetic nanoparticles, (b) chitosan, and (c) CS-MNPs.

*3.5 Zeta Potential* In order to further prove the influence of chitosan polymers on the property of $Fe_3O_4$ nanoparticles, zeta potentials of bare MNPs and CS-MNPs were collected at different pH values (Fig. 2(b)). The zeta potentials, as a measurement of repulsive or attractive forces between particles, play an important role to find not only the isoelectric point (pI), but also the colloidal dispersion stability. Zeta potentials of uncoated and chitosan polymer coated magnetic nanoparticles (0.1 mg·L$^{-1}$) were measured in $10^3$ M NaCl aqueous solution at different pH values. The solution pH was adjusted by NaOH or HCl. Results show that PI values of unmodified and CS modified MNPs are 6.7 and 4.8, respectively. Upon surface modification with CS polymer containing multiple carboxyl groups, the PI shifts to a lower pH value because of the introduction of acidic surface groups.

*3.6 Heating ability of CS-MNPs*

The induction heating instrument used for calorimetric measurement is shown in figure 1(e). The temperature vs. heating time for aqueous suspension of chitosan-coated iron-oxide nanoparticles (50 mg ml$^{-1}$) exposed to the alternating magnetic field with different intensities (H = 10 and 20 kA m$^{-1}$, f = 100 kHz) are shown in figure 1. As shown in figure 1(e), after a short time (less than 3 min), the temperature of the suspension is saturated at $T_{max}$ = 46 and 59°C for H = 10 and 20 kA m$^{-1}$, respectively. This saturation is due to the decreasing of the magnetization near the $T_c$ and the balance of heat exchange with surrounding medium. Also, the heating curves show that the maximum temperature of the magnetic suspension falls within the desired temperature range (40-60°C) which suggests that the prepared samples are suitable for biomedical applications.

## 4. Conclusions

Magnetic nanoparticles were successfully synthesized by the co-precipitation method. This synthesis technique resulted in nanoparticles with an average size of 16 nm. The pure obtained nanoparticles were then successfully encapsulated with 4-nm-thick chitosan coating. The formation of chitosan on the surface of nanoparticles was confirmed by XRD, FTIR and SEM analyses. Cytotoxicity of bare and chitosan-coated nanoparticles was evaluated by MTT assay with MCF-10A and MCF-7 cell lines (data not shown). The results revealed that the both samples have negligible toxicity (below 500 μg ml$^{-1}$). Heating experiments at safe magnetic field (f=100 kHz, H=10-20 kA m$^{-1}$) revealed that the maximum achieved temperature of water stable chitosan-coated nanoparticles (50 mg ml$^{-1}$) is fully in agreement with cancer therapy and biomedical applications.

**Acknowledgments:** The research leading to these results has received funding from the European Union Seventh Framework Programme (FP7-PEOPLE-2013-COFUND) under grant agreement n° 609020 - Scientia Fellows.